# Models of the Hydrodynamic Histories of post-AGB Stars.
# I. Multiflow Shaping of OH231.8+04.2

Bruce Balick[1], Adam Frank[2], Baowei Liu[2], & Martín Huarte-Espinosa[3]

[1] Department of Astronomy, University of Washington, Seattle, WA 98195-1580, USA; balick@uw.edu

[2] Department of Physics and Astronomy, University of Rochester, Rochester, NY 14627, USA; afrank@pas.rochester.edu, baowei.liu@rochester.edu

[3] Center for Advanced Comp & Data Systems, University of Houston, 4718 Calhoun Rd., Houston TX 77204-3058, mhuartee@central.uh.edu

ABSTRACT

We present a detailed hydrodynamic model that matches the present structure of the well-observed preplanetary nebula ("pPN") OH231.8+04.2 ("OH231"). The purpose of the model is to present a physically justified and coherent picture of its evolutionary history from about 100 years of the start of the formation of its complex outer structures to the present. We have adopted a set of initial conditions that are heavily constrained by high-quality observations of its present structure and kinematics. The shaping of the nebula occurs while the densities of the flows are "light": less than the surrounding AGB-wind environment. The simulations show that pairs of essentially coeval clumps and sprays of the same extent and density but different outflow speeds sculpted both the pair of thin axial flow "or spine" and the bulbs. The total ejected mass and momentum in the best fit model are surprisingly large—3 $M_\odot$ and $2.2 \times 10^{41}$ gm cm s$^{-1}$, respectively—however, these values are reduced by up to a factor of ten in other models that fit the data almost as well. Our ultimate goal is to combine the present model results of masses, momenta, flow speeds, and flow geometries for OH231 with those of other models to be published in the future in order to find common attributes of their ejection histories.

Subject headings: planetary nebulae: individual (OH231.8+04.2); individual stars: AGB and post-AGB; stars: winds, outflows

1. INTRODUCTION

The traditional picture of AGB winds and the formation of proto planetary nebulae ("pPNe") is that they are driven by isotropic radiation pressure absorbed on dust particles that condensed in expanding and cooling quasi-isotropic winds. It's clear that in most cases reality is far more interesting, albeit still only partially understood. Dozens of optical images of pPNe from the Hubble Space Telescope[1] ("HST") as well as images in molecular emission lines and dust-emitted continuum light from millimeter-wave interferometers such as ALMA show that the wind streamlines are far from isotropic. Other studies (e.g. Bujarrabal et al., 2001A&A...377..868B, hereafter "B+01") have demonstrated that the momentum in the outflows is often far too large to be explained by the winds of a single AGB star. Moreover, the predominant kinematic pattern of pPNe is one in which Doppler shifts that increase steadily with radius from the star—in direct conflict with expectations of a steady jets from stellar outflows.

---

[1] Observations were made with the NASA/ESA Hubble Space Telescope and downloaded from the Hubble Legacy Archive, which is collaboration between the Space Telescope Science Institute (STScI/NASA), the Space Telescope European Coordinating Facility (ST-ECF/ESA), and the Canadian Astronomy Data Centre (CADC/NRC/CSA). The Space Telescope Science Institute is operated by the Association of Universities for Research in Astronomy, Inc., under NASA contract NAS 5-26555.

Observing the active flow acceleration and collimation process presently lies far beyond our grasp. The challenges are that we are generally unable to catch the brief ejection event at its start or to monitor its subsequent evolution using telescopes that spatially resolve the structure of the outflow in its infancy as it grows into its final form. So the nature of the ejection and collimation processes must be inferred. The most plausible general explanations for the various types of bipolar pPNe—and those for which there is slowly mounting evidence—is that pPNe are formed by mass ejected in magnetically shaped winds from the active surfaces of AGB stars or from disks or mass overflows associated with mass transfers in close binary systems (Balick & Frank 2002ARA&A..40..439B, Nordhaus et al., 2007MNRAS.376..599N, De Marco 2009 PASP..121..316D, Soker & Kashi 2012ApJ...746..100S, Chen et al. 2016MNRAS.460. 4182C, Staff et al. 2016MNRAS.458..832S, Iaconi et al. 2017MNRAS.464.4028I).

In this paper we present hydrodynamical models that successfully fit extant observations of one pPNe, an appropriately well-observed pPN OH231.8+04.2 (hereafter "OH231"). However, as them models will show, the observably expanding structures are presently coasting long after the completion of their active shaping process. Thus in order to surmise its evolution backwards from the present, we must use hydrodynamical models of two pairs of axisymmetric flows to provide insight into the early ejection process. OH231 is one of the most advantageous of its class for modeling owing to its large angular size, well-resolved and regular shape, and a rich literature of observational data.

In particular we will show that the models yield a simple and robust set of tightly constrained initial values for the hydrodynamic state variables that describe the evolution of OH231 and the mass and momentum of the outflows that have shaped it. The data that we used to constrain our hydrodynamic simulations will be discussed in section 2. In section 3 we describe our methodology and the starting point for the simulations. The results of the models are presented in section 4. In section 5 we summarize the results and their relevance for understanding the histories of pPNe.

Riera et al. (2005RMxAA..41..147R; hereafter "RRA05") constructed a hydro model for OH231 consisting of a pair of non-axisymmetric and brief columnar jets, or "jet-like pulses", injected at 200 km s$^{-1}$. The northern pulse passed close to a companion AGB star with a continuous asymmetric wind that is located near the end of the lobe. The interaction of the pulse with the outer parts of the AGB wind impeded its progress. We will compare the two sets of results later in the paper.

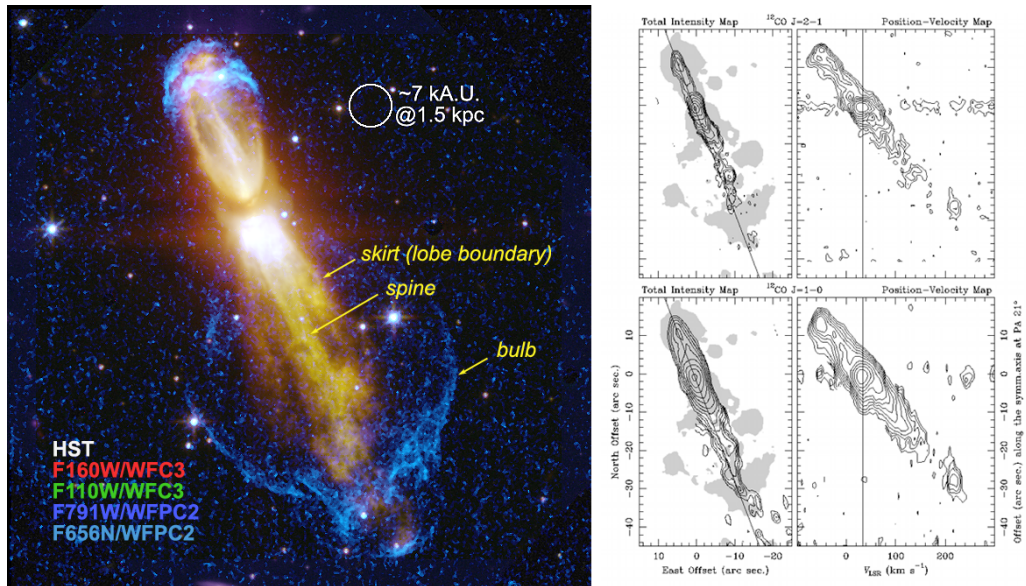

Fig 1. Images of OH231 in optical light (left) and $^{12}$CO (right) taken from the literature. Left: An overlay of optical and near IR images of OH231 obtained from the Hubble Legacy Archive. North is up. The white circle provides an image scale. Right: The spatial distribution and axial Doppler speeds of $^{12}$CO lines from OH231 obtained at the Plateau do Bure interferometer and republished with permission[2]. The upper (lower) row shows images and kinematic results for the $^{12}$CO(1–0) ($^{12}$CO(2-1)) line.

## 2. RELEVANT OBSERVATIONS

We summarize the observations of OH231 that are used to contract and constrain the present model simulations; to wit, the current spatial structure of the hydro state variables density, vector momentum, and temperature. The associated proxy observables are the brightness distribution of the images, patterns of Doppler shifts (Fig. 1) and the $^{12}$CO kinetic temperature. In the discussion that follows, and unless otherwise noted, we summarize the characteristics of this iconic and relatively massive pPN compiled from papers by Kastner et al. (1992ApJ...398..552K), Sánchez Contreras et al. (1997A&A...327..689S, 2000A&A...355.1103S, hereafter "S+00A", 2000A&A...357..651S, hereafter "S+00B"), Alcolea et al. (2001A&A...373..932A, hereafter "A+01"), and Bujarrabal et al. (2002A&A...389..271B, hereafter "B+02")[3].

*Geometry.* OH231 consists of two opposite flame-shaped "lobes" similar shapes (Fig. 1, yellow and white colors) that are enclosed within bulbous structures called "bulbs". Of vital importance is the high resolution optical and near-IR images of OH231 obtained using the Hubble Space Telescope ("HST")[4]. All of the extended structures are outlined in shock-excited lines (Fig. 1, blue colors) (S+00A&B). The leading tip of the northern (southern) lobe is 18″ (38″) from the dark lane that separates them. The lengths of the $^{12}$CO outflow lobes are slightly smaller. The distance to OH231 is 1.5 kpc, so the tip of the north (south) lobe lies at 32,000 (66,000) A.U. from the nebular core after correction for the nebular inclination of 35° (a factor of 1.22).

---

[2] CO panel source: a web document entitled "From the AGB to the PN phase with the Plateau de Bure Interferometer" posted as a pdf file by A. Castro-Carrizo with J.M. Winters and R. Neri, c. 1998, with no public URL. These CO results were subsequently reformatted and republished by A+01.
[3] It should be noted that Table 1 of B+02 lists many local properties of very specific nebular structures that are measured using various emission lines. We do not try to match their results in detail since our models are intended primarily to mimic the large-scale structure of OH231.

The lobes and bulbs extend from a dense, compact, highly structured, slowly expanding and only partially resolved central core. All of them share a common symmetry axis. In addition A+01 make a compelling argument that the core, observable features of OH231 are surrounded by an isotropic, stratified, and low-surface-brightness density distribution consistent with a wind from a central AGB star whose origin preceded the ejection of the lobes and bulbs by thousands of years. See A+01 (section 3.2.1) for a discussion of the history of the AGB winds. It is commonly assumed that the AGB wind is steady. If so then its radial density profile declines as $r^{-2}$.

*Aggregate properties.* The total $^{12}$CO mass of OH231 is 1 $M_\odot$ (A+01), one-third of which is in the two spines (the fast flows) that lie along the nebular symmetry axis (Fig.1) and rest is slow gas near the core. The remainder lies in the slow AGB wind and core. The shorter and slower northern spine has slightly more mass and momentum (<15%) than its southern counterpart (A+01). Given their similar masses but differences in their speeds, the slower, brighter northern spine presumably contains about twice the momentum as its southern sibling.

The total linear momentum of OH231 seen in $^{12}$CO, ≈27 $M_\odot$ km s$^{-1}$, is about $10^5$ times larger than the linear momentum that the central star of OH231 can supply (B+01). Thus radiation pressure supplies a trivial fraction of the momenta of that formed the surrounding structures. Such large momentum "surpluses" are rare among pPNe, but not unprecedented.

*Bulbs.* Bulbs surround both the north and south lobes whose perimeters. The southern lobe is by far the most conspicuous, having dimensions of roughly 27″×30″ (40,000 × 45,000 A.U.). The vertex at the base of the bulbs is about 4000 A.U. in width. The inner half of the southern bulb has straight edges that open at angles of ~ ±50° before they re-converge at the protruding tip of the southern lobe. This lobe shows signs of ripple-like instabilities along its forward edge. Hα lines inside the lobe exhibit a trend of monotonically increasing Doppler shift with distance from the core with speeds as large as 400 km s$^{-1}$. A high-contrast rendition of the Hα image shows that the northern bulb is thinner than the southern bulb. Its aspect ratio is about 1:2 and its opening angle is about 40° at the base.

*Lobes.* The lobes are flame-shaped in outline. (This morphology is common among the lobes of pPNe.) Each lobe is comprised of a spine plus a skirt (Fig. 1). The brighter northern lobe has a rounded leading edge within which a knot of HCO$^+$ (a tracer of high-density gas ≳ $10^6$ cm$^{-3}$) is located. In contrast, there is no prominent peak in any molecular tracer or starlight-scattering dust near the end of the southern lobe. Rather the spine slowly fades and becomes irregular starting about halfway from the core.

---

[4] The HST images in the F656N and F791W filters shown in Fig. 1 are derivatives of images obtained in Dec. 2000 and originally published by B+02. The F110W and F160W images were obtained Balick et al. in Oct. 2009 (GO12600). All of these images were downloaded from the Hubble Legacy Archive.

Forde & Gledhill (2012MNRAS.421L..49F) found patchy areas of $H_2$ emission near the leading edge of the northern lobe. $H_2$ is typically emitted in shocks of speeds of a few tens of km s$^{-1}$. Thus the $H_2$ lines may arise behind the lobe tips where putative outflows overtake and shock the leading edge of the bulb at modest speeds (20–50 km s$^{-1}$).

*Spines.*   Thin, bright spines run down the nebular symmetry axis in both lobes (Fig. 1). Spines have a central relevance here since, of all of the features of OH231, they alone have clearly defined kinematic patterns. The overall shapes of both spines are similar. $^{12}$CO within each lobe is intimately associated with its spine (but not the skirts). $^{12}$CO observations show that the tip of the southern (northern) spine has an inclination-corrected outflow speed of about –400 (+200) km s$^{-1}$. The outflow Doppler shift of the spines rises linearly with distance. This kinematic pattern is sometimes called a "Hubble-like flow pattern" that has been interpreted by A+01 and RRA05 as a sign of a ballistic jet-like outflow. (We will offer a very different alternative interpretation later.) The spines share the same speed gradient and, thus, the same kinematic ages, ≈770y.

*Southern Skirt.*   The skirt (Fig. 1) is seen only in reflected starlight. So its density, kinematics, and mass are not directly measureable. The wedge-shaped lateral edges of the skirt suggest that the skirt may be starlight that is scattered from dust lying within in a stellar illumination cone of opening angle ±20˚. More speculatively, the skirt may be a dusty conical outflow; however, no swept up gas is found along its outer edge.

In addition the central stars of OH231 and their possible roles in the ejection of the nebular features of OH231 are discussed by Sánchez Contreras, Gil de Paz,& Sahai 2004ApJ...616..519S. Of particular relevance here is that the AGB star, QX Pup, has an A-type companion. The binary originally has a total mass in excess of 4.5 $M_\odot$. This is an upper limit on the nebular mass today. They also argue that the momentum of the nebula is supplied by winds from an accretion disk the companion with an orbital radius ≲50 A.U. The ejection of the gas took place within 100 yr of its onset. The mass loss rate needed to supply the fast outflows was of order $10^{-3}$ $M_\odot$ yr$^{-1}$.

3. METHODOLOGY & PARADIGMS

3.1 Methodology

The code AstroBEAR (Cunningham et al., 2005ApJ...631.1010C, 2009ApJS..182..519C) was used to construct hydrodynamic models of OH231. AstroBEAR solves the Eulerian equations of fluid dynamics on a 3-D grid using high-performance adaptive mesh refinement as needed to resolve shapes and shocks. The kernel of AstroBEAR is a versatile 3-D hydrodynamic solver that can be applied in a wide variety of circumstances, including magnetized/cooling flows and rotating coordinating systems (Huarte-Espinosa et al., 2012ApJ...757...66H). The code does not contain facilities for simulating emission-line images, predicting an emergent spectrum, or for following the transfer of radiation.

The numerical code accommodates three types, or paradigms, of mass injection from a central source into the ambient gas: clumps (a spherical knot of a single ejection age), cylindrical jets (hereafter "jets": a thin flow consisting of parallel streamlines ejected at constant speed), and steady tapered conical flows or "sprays" (diverging streamlines as described later). Each of these can be launched into the AGB environment in any order. In all models we assume that the ambient gas consists of AGB ejecta and that the flows are axisymmetric and share the same symmetry axis and a common origin (Fig. 1).

We performed the computations in one quadrant of the 64 × 256 2-D model grid (32,000 × 128,000 A.U.) in which the injected gas is launched into a stratified ambient medium along the $y$-axis at time t = 0. However, computational speed is important since multiple free parameters must be adjusted for a successful fit. Thus we did not attempt 3-D because the amount of computing time is prohibitive. The computational complexity of 3-D models is only needed to follow the growth patterns of small-scale instabilities at shock interfaces. This is not our primary interest.

The basic grid size, 500 A.U., was decreased by the adaptive mesh algorithm by up to five factors of powers of two in regions where local pressures have steep gradients. Given the number of free parameters we ran over 60 different models at modest resolution to find a good low-resolution outcome. Subsequently we reran the simulation enhancing the resolution by one or two additional factors of 2. Even so, these "full" models often required 24 hours to execute. Thus, after the flow configuration had grown substantially and attained its final form (after a few hundred years) the resolution was relaxed by a factor of two with no noticeable loss of overall morphology.

3.2 Flow Concepts

We consider the most appropriate flow concepts for forming the bulbs and the spines seen in HST images. The bulbs are formed by a wide-angle, geometrically tapered polar flow called a "spray" with an opening angle consistent with the HST images (see below). Narrower flows, such as jets and clumps, were considered for the formation of the spines. The age of OH231 is rounded to 800 yr.

*Ambient Gas.* We assumed that the ambient AGB mass deposited on the grid prior to $t = 0$ is isotropic with a density distribution $n_{amb}(r) = n_{amb,0} (r_0/r)^2$. Here $n_{amb,0}$ is the density of the ambient medium at a fiducial radius $r_0$. We adopted adopt $n_{amb,0} = 2 \times 10^4$ cm$^{-3}$ at $r_0 = 3500$ A.U. The total mass of the ambient gas beyond 1000A.U. is ~0.7 M$_\odot$, in accord with the estimate of A+01. Our simulations are restricted to a 2-D slice through this density distribution.

*Bulbs.* The bulbs are presumably the outcomes of conical nuclear sprays that have displaced the ambient gas within their volumes. The spray exits a nozzle at an opening angle $\Theta_{spr}$ with density $n_{spr}$, speed $v_{spr}$, and kinetic temperature $T_{spr.}$ The density and speed are modulated in polar angle by a Gaussian of 1/e width $\phi_{spr}$, where $\phi_{spr} \approx \Theta_{spr}$. This modulation of the flow momentum "softens" the edges of the flow. (In the absence of this modulation, a conical flow forms an open wedge-shaped lobe with straight sides when injected into an ambient medium with the same radial density distribution.) Lee & Sahai 2003ApJ...586..319L (hereafter "LS03") showed that some sprays form closed flame-shaped lobes.

*Spines.* The spines are relatively thin axial features whose shapes and kinematics are defined by $^{12}$CO observations that were presented by A+01. The spines exhibit a linear kinematic gradient seen that is fundamentally inconsistent with any type of steady collimated flow such as a jet. For this reason we consider that the spines are similar to the wakes of clumps ejected in the same event as the sprays.

The clumps are characterized by their properties at the nozzle upon ejection at $t = 0$: an initial radius $r_{cl}$, initial density $n_{cl}$, ejection speed $v_{cl}$, and temperature $T_{cl}$. These are free parameters though their starting values are extracted from the observations of section 2. The mass of the clump at launch is $1.54 \times 10^{-8}$ ($n_{cl}$/cm$^{-3}$) ($r_{cl}/10^3$ A.U.)$^3$. A+01 estimated that the total mass of high-speed gas (i.e., the spines) is on the order of 0.2 M. We also adopt $T_{cl} = 10$K; this value is of no consequence to the outcomes so long as $T_{cl} < 10^5$K.

*Placement of the nozzles.* The bulb must be launched into the lee of the spray in order that the wake that follows the clump remains intact. If their locations are reversed then a spray of appropriate total momentum flux to account for the bulbs will destroy the structures in the lee of the clump and eradicate the spine. The details are discussed within the context of specific models.

A note about alternate paradigms: A+01, Meakin et al. (2003ApJ...585..482M), RRA05, and others have speculated that cylindrical jets are the most viable paradigm provided that the jets are accelerated *in situ* in such a way to emulate a ballistic jet. The *in-situ* acceleration mechanism is unspecified and we are not able to imagine what it might be.

4. RESULTS

The best-fit simulation of the flow characteristics of OH231 is shown in Fig.2 panel A. Table 1 lists the initial conditions that were used for the model. The major features of the model nicely match the observations, including the speed gradient observed along both spines.

The densities of the clumps and sprays at the launch surface $n_{cl,0}$ and $n_{spr,0}$, their launch speeds $v_{cl,0}$ and $v_{spr,0}$, the radii of the clumps $r_{cl}$, the nozzle widths, geometries and durations of the sprays, $r_{spr}$, $\Theta_{spr}$ and $\phi_{spr}$, are, in principle, free parameters. Their initial values were estimated as follows: $v_{cl,0}$ and $v_{spr,0}$ were set equal to the observed speeds at the end of the spines: 225 (400) km s$^{-1}$ to the north (south). The clump radius was set to the size of the circle in Fig. 1; $r_{cl} = 3500$ A.U. $\Theta_{spr}$ and $\phi_{spr}$ were directly measured from the H$\alpha$ images of OH231 at the base of the flow and set equal to one another. The densities $n_{cl,0}$ and $n_{spr,0}$, were found by trial and error for each lobe.

The northern and southern sprays and clumps are launched at $t = 0$ from two different spherical nozzles that are aligned on the *y* axis. The nozzle for each spray has a radius of 5000 A.U so that the spray initially envelops the clump on the grid (see section 3.2) and precedes it into the ambient gas (until the clump overtakes the leading edge of the spray after several hundred years). After 800 yr we estimate from the initial conditions that the total mass (momentum) of gas injected into the spray and clump is 2.8 M$_\odot$ (~1000 M$_\odot$ km s$^{-1}$ = $2 \times 10^{41}$ g cm s$^{-1}$).

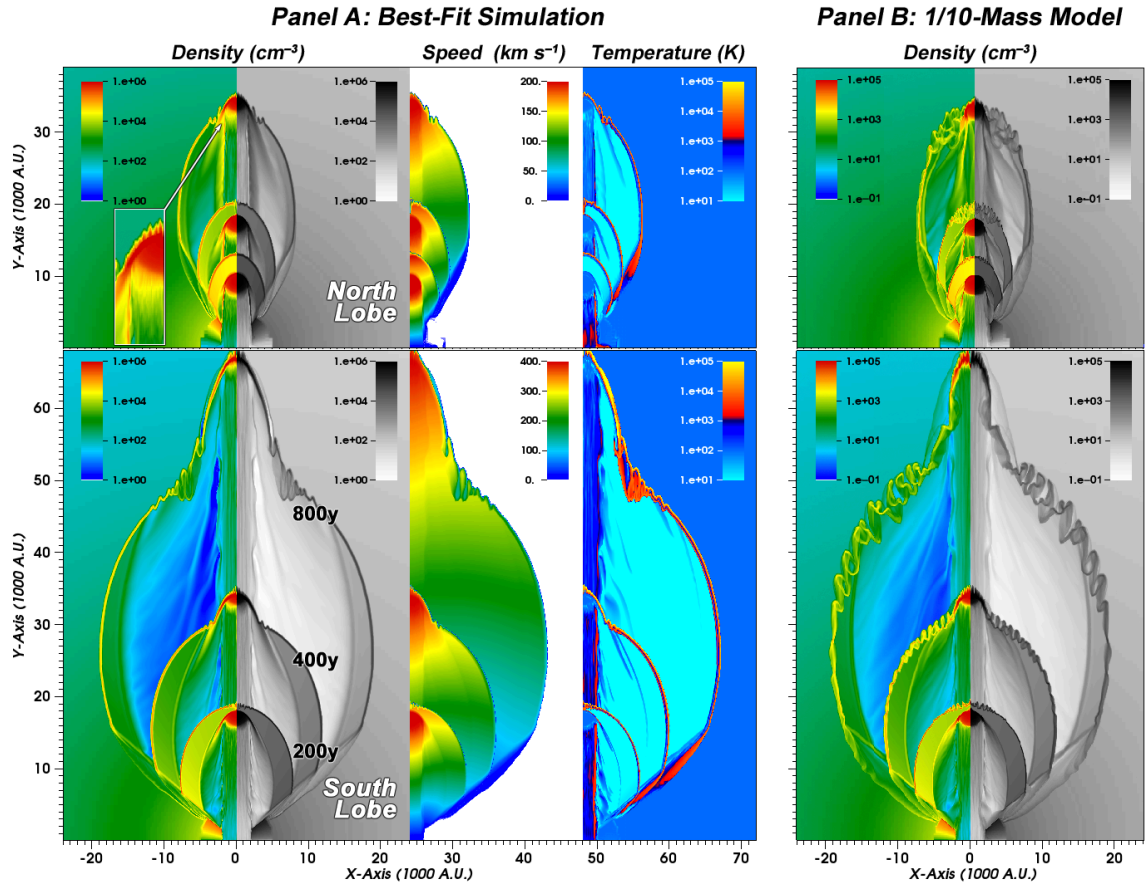

Fig 2. Panel A: Best-fitting simulations of OH231 after 200, 400, and 800 yr after the clump and spray are launched. Two left panels: two complementary representations of the model density distribution. Two right panels: the radial flow speed and temperature. Parameters of the model are found in Table 1. The inset shows the trail of gas being shed into the spine as the clump crushes. It applies also to the southern clump at $t \approx 200$ yr. Note that the color bars for the speed panels differ for the north and south lobes. Panel B: Like A for the density distribution of a simulation where all of the initial densities (and the total mass) are reduced by a factor of 10. It demonstrates that the outcomes of isoflow models can have similar large-scale outcomes.

Table 1. Initial Parameters for the Best Model

| Model Feature | Value | | Note |
|---|---|---|---|
| **Ambient (Pre-ejection) Gas** | | | |
| density $n_{amb,0}$ (cm$^{-3}$) | $2 \times 10^4$ | | a |
| fiducial radius ($r_o$) (A.U.) | 3500 | | b |
| speed $v_{amb}$ (km s$^{-1}$) | 0 | | c |
| temperature $T_{amb}$ (K) | 10 | | b |
| total mass on grid (M$_\odot$) | 0.7 | | b |
| **Clump at launch ($t = 0$)** | North | South | |
| initial radius $r_{cl}$ (A.U.) | 2000 | 2000 | c |
| initial density $n_{cl,0}$ (cm$^{-3}$) | $10^6$ | $10^6$ | a |
| ejection speed $v_{cl,0}$ (km s$^{-1}$) | 225 | 400 | b |
| initial mass (M$_\odot$) | 0.1 | 0.1 | b |
| initial temperature $T_{cl}$ (K) | 10 | 10 | a |
| total injected momentum (gm cm s$^{-1}$) | $4.5 \times 10^{39}$ | $8.0 \times 10^{39}$ | b |
| **Spray at the nozzle surface** | North | South | |
| radius $r_{spr}$ (A.U.) | 5000 [c] | 5000 | b |
| density $n_{spr,0}$ (cm$^{-3}$) | $7 \times 10^5$ | $7 \times 10$ | c |
| ejection speed $v_{spr,0}$ (km s$^{-1}$) | 225 [b] | 400 | b |
| temperature $T_{cl,0}$ (K) | 10 [a] | 10 | a |
| polar opening angle $\Theta_{spr}$ | 30° | 50 | b |
| Gaussian 1/e taper angle $\phi_{spr}$ | 30° | 50 | b |
| mass injection rate $\dot{M}_{spr}$ (M$_\odot$ y$^{-1}$) | 0.0048 | 0.023 | c |
| injection duration $\Delta t$ (y) | 100 | 100 | c |
| total injected mass (M$_\odot$) | 0.48 | 2.3 | c |
| total injected momentum (gm cm s$^{-1}$) | $1.8 \times 10^{41}$ [c] | $2.2 \times 10^{40}$ [c] | c |

Notes:
a the shapes and sizes of the model outcomes are not sensitive to the adopted value
b the value is constrained by observations
c the mass is not directly constrained by observations

The "best model" in Fig 2 panel A is simply one of a family of very similar simulations that nicely match the shapes and kinematics of OH231 (Fig. 1). Other members of the family, or "isoflows", share the same initial parameter values except for their densities. However, the ratios of all densities, $n_{amb}$, $n_{spr}$, and $n_{cl}$, are the same for all of the isoflows. The "best model" is simply the isoflow that satisfies the $^{12}$CO mass requirements of OH231: mass of slow (fast) gas ≈ 0.7 M$_\odot$ (0.2 M$_\odot$). The bulb is not detectable in $^{12}$CO.

However, the masses and momenta of the flows that shape the spine and bulbs in the best-fit model seem excessive (~factor of 10–100) relative to their values in other pPNe (B+01, section 2). Specifically, B+01 estimate a combined momentum for the spines of OH231 of 4 x 10$^{39}$ g cm s$^{-1}$ (with very large uncertainties). Our result for the spines agrees reasonably well with this value. The momentum of the bulbs dominates the combined total. This momentum is plausible because the volumes of the spray-formed bulbs are about 100 times larger than the clump-formed spines. Thus the sprays have displaced up to 100 times more ambient mass (and momentum) than the spines.

For comparison, the total mass and momentum of the isoflow shown in Fig. 2 Panel B is ten times smaller than the best-fit model even though both simulations have the essentially the same shapes, dimensions, and kinematics.

4.3  Comparison to key observations.

The outcome of a successful hydro model must replicate the fundamental features of OH231, including the shapes, dimensions, and overall geometry of the spines and bulbs (Fig. 1). The speeds of the gas must exhibit matching linear gradients from the core to the lobes' tips with a terminal speed that matches the $^{12}$CO observations. A dense ($\geq 10^6$ cm$^{-3}$) knot-like feature that accounts for the presence of HCO$^+$—a tracer of high-density cold gas—must lie at the leading edge of the northern lobe but not the southern.

*The spray-formed bulbs.*  The leading edge of ambient gas that is displaced by the spray forms the visible, shocked outer edges of the bulb. The leading edge of this surface quickly displaces about its own mass of ambient gas and slows down. As a result the clump eventually overtakes the bulb edge and forms polar protrusions in each lobe.

After 800 yr nearly all of the mass in the original spray has been incorporated into the high-latitude walls of the bulb (Fig. 2). At mid latitudes a fraction of the initial spray still resides in a thin outer crescent. The lobes become bulbous since the walls at these lower latitudes are slower and contain relatively less mass than at higher latitudes. Moreover the streamlines strike the reverse shock at mid-latitudes obliquely, so the shock speed is about half as large. For these reasons the walls at mid latitudes are barely visible in shock-excited tracers. Similay the Doppler speeds of the mass within the walls will decrease with latitude, as observed. Other regions inside the lobes and outside the spine are essentially empty.

Note that an alternate way to form a bulb is by thermal rather than ram pressure. We explored the injection of a hot, thermally over-pressured low-mass wind ($\approx 10^6$ K) of injection density ($\approx 10^5$ cm$^{-3}$) and small flow speed (<100 km s$^{-1}$). However, all such hot-wind simulations failed to recreate the smooth bulb edges. The thermal overpressure behind the walls creates unstable, persistently and strongly serrated walls as soon as the hot gas is suddenly released into the stationary ambient gas.

*The clump-formed spines.*  Each clump has an initial density of $10^6$ cm$^{-3}$, a radius of 2000 A.U., and a mass of 0.1 M$_\odot$. The clumps are launched into the north (south) lobe at a speed of 225 km s$^{-1}$ (400 km s$^{-1}$), as implied by the deprojected Doppler speeds of the $^{12}$CO at the flow tips. The combined momenta of the clumps are 60 M$_\odot$ km s$^{-1}$, or twice the uncertain estimate of A+01.

We emphasize that the clumps are identical aside from their launch speeds. Although the clumps slow down slightly in transit, their terminal speeds remain on the order of their launch speeds as they flow through the co-moving streamlines of the spray. At first the clumps only gently shock the almost co-moving gas through which they flow. By $t = 800$ yr each clump has overtaken the leading edge of its bulb.

At $t = 800$ yr the remnant of the southern clump consists of a thin crescent-shaped leading rim of compressed gas with density $n_{cl} \approx 4 \times 10^6$ cm$^{-3}$. Within its trailing spinal column the speed of the gas rises linearly from zero at the clump's launch origin to 400 km s$^{-1}$ at the flow tip at all times (until the clump crushes shortly after $t$ = 800 yr). The spine's density, $n_{spine,}$, rises from 50 cm$^{-3}$ at $y$ = 5000 A.U. to 3000 cm$^{-3}$ behind the remnant of the clump. The trailing half of the northern clump is still intact at $t$ = 800 yr. The thin rim of compressed clump gas at its leading edge has a density $n_{cl} \approx 4 \times 10^6$ cm$^{-3}$.

A clump crushes when it transfers enough momentum to the ambient gas that it significantly slows down and, in doing, so, initiates internal supersonic shear and turbulence. The crushing time, $t_{crush} = (2r_0 / v_{cl}) \cdot (n_{cl} / n_{amb})^{1/2}$. In our best simulations neither the south nor the north clump crushes in 800 yr, though the south clump is on the verge when the simulation ends. Note that the onset of crushing of the clumps—especially the northern one—is appreciably delayed since it spends most of its time immersed in the co-moving spray. Note also that the crushing time is the same for all isoflows.

The spines develop immediately in the stagnation column behind the clump. The dominant important source of gas in the spine is gas that was stripped from the outer edges of the clump in transit and falls behind (inset, Fig. 2). The speed gradient within the spine develops very quickly and persists through 800 yr (before the clump crushes).

Thin layers of hot, shocked ambient gas precede the leading edge of both clumps' remnants into the ambient gas after the clumps penetrate their respective bubble edges. This rim hot, shocked gas expands thermally and, for the most part, laterally. It is left behind as the remnants of the clump move onward. Appropriate emission lines will be excited behind the attendant shocks. Although ambient gas does not flow through the local contact discontinuity ("CD"), the CD moves steadily into the ambient medium.

At the end of the simulation the dense gas at the tip the northern clump can be seen in high-density tracers such as HCO$^+$ (see section 2). The southern clump is almost crushed and its remaining mass and average density are dropping. It might not be detectable in the isoflow with 10-times-lower density mentioned above.

*Skirts.* The simulations fail to reproduce the skirts in the southern lobe that are readily seen in all broadband HST images of OH231. We speculate that the skirts consist of forward-scattered light from dust in the bulb that lie within a stellar illumination cone of opening angle $\approx \pm 20°$.

4.4  Parameter Ranges and Model Robustness

The model discussed above is the "best" of about five-dozen models. In the process of exploring the values of the initial conditions we were able to understand how changes in the starting parameters affect the final outcomes. Here we summarize our experiences.

In our parameter variation studies we almost always adopted the clump flow speeds and the static ambient density distribution and varied other parameters. We find that changes in $n_{amb}(r_o)$ by factors of about three (with no other changes) clearly had an impact on the evolution of the spray since the spray displaces a considerable volume of static ambient gas and slows down. In short, for a fixed speed of the spray, bulbs of the appropriate dimensions are formed only when that ratio of $n_{spr}(r_o)$ and $n_{amb}(r_o)$ is within 10% of 35. This is a robust result that applies to both lobes.

The outcomes are similar for nearly all values of the clump to spray density ratio so long as the clump is at least as dense as the spray at launch. (However, if $n_{cl}(r_o)/n_{spr}(r_o) \leq 3$ then the clump or its crushed remains cannot overtake the outer edge of the expanding bulb within 800 yr.)

Given the adopted ambient medium, the launch speeds of both flows could not be varied by more than a few percent without a very clear impact on the lengths of the lobes. The requirement that the speed gradients are the same in the spines of both lobes very firmly requires that the launch speeds of the clumps must be in exact proportion to their final offsets from the nucleus; in this case ~2:1.

The opening angle of the sprays determine the aspect ratios of the bulbs at 800 yr. Departures from our adopted values by 10° or more have noticeable effects on the final dimensions of the bulbs. The same applies to the polar width of the Gaussian taper that we applied to the density and speed of the conical flow to give the spray its overall form.

The large initial radius of the clumps that we have adopted, 2000 A.U., is obviously an issue given that the dimensions of a binary star system or the dominance of a strong surface magnetic field, 100 µG, will be far below 100 A.U. We have adopted this clump radius for a practical reason—the clump results in an axial protrusion of the proper total width, ~7000 A.U.—and an expedient one—we easily resolve it on the model grid. We speculate that, with suitably adjusted initial conditions, the "magnetic bomb" model of Huarte-Espinosa et al. (2012) might produce clumps of the appropriate size and speed within the first hundred years after launch. In any event, readers are cautioned that the clump might simply be a region born with complex internal structure and dynamics.

4.5  Comparison with RRA05 model

RRA05 approached the hydrodynamical modeling of OH231 differently than did we. They launched columnar pulses of various properties from the central star into asymmetric ambient media in order to account for the geometric properties of the lobes. The dense axial cores of the pulses have a radius of 3300 A.U. (about the same as the size of the clump in our best-fit simulations) and a core pulse density of $10^4$ cm$^{-3}$ (100 times sparser than the best-fit clump in our simulations). The ejection speed within each pulse decreases linearly with distance from the source from 200 km s$^{-1}$ at their leading edge, so the Hubble-like flow pattern that develops as each pulse pushes forward is forced by the initial conditions. (In our simulations this gradient is a natural, robust, and persistent consequence of the clump-AGB-wind interaction during the early light-flow phase.)

One pulse flows through the winds of an AGB star located near the end of the northern lobe and offset by various amounts from the flow axis. (NB: No AGB star or evidence of a reddened disk appears in about the proper location in an HST image at 1.6 µm.) The AGB wind has an equatorial density contrast $\xi$ that varies from 1 to 50 in their five simulations, and its equator is orthogonal to the pulse flow axis. In models with $\xi \geq 25$ the pulse is impeded as travels through the edge of the wind disk and slows down. (In effect, the disk of the offset AGB star serves as a sheet-like barrier through which the pulse penetrates.) The lobes created by pulses of disparate initial characteristics successfully develop the observed ratio of the lengths of both lobes and reproduce their speed gradients. However, the ratio of the Doppler shifts at the end of the lobes is close to unity and the lobes have no spines or skirts.

5 SUMMARY AND CONCLUSIONS

We have constructed a successful hydrodynamical simulation of the lobes of OH231 using a combination of pairs of clumps that are embedded within sprays of brief duration and launched into static ancient AGB winds. (This is not unlike the ejection of a cork and a spray of high-pressure fluid when a bottle of champagne is opened, except that in our case the spray is ejected shortly before the cork.) After 800 yr each of the clumps leaves behind a tail, or spine, in which the steady kinematic gradient seen in the Doppler shifts of $^{12}$CO is replicated. The brief spray with an opening angle of ±50° expands to form the wide southern bulbs. Its northern counterpart is smaller, denser, and thinner since it is ejected at about half the speed and opening angle.

The values of the initial conditions adopted for our best simulation are almost fully constrained with a wide variety of observations of the structure, kinematics, and molecular masses of OH231 (Fig. 1). In addition to matching the shapes of the lobes, one feature of the models is especially conspicuous: the linear speed gradients along the outflows. The injected gas is characterized by a single speed in each lobe of OH231 and lies at the lobe perimeter as the simulation evolves. So the speed gradient is not imposed by initial conditions. It is a natural and robust result of a combination of swept up ambient gas and clump material that has been shed from the edges of the clumps. Steady jets or sprays of any geometry are incapable of explaining these speed gradients. In addition, they would not replicate the morphology of the bulbs while also producing the highly conspicuous spines.

We also showed that the same outcomes of morphologies and kinematics are expected over a range of injected masses provided that the ratios of the densities of the ambient gas, the clumps, and the spray are fixed. That is, members of the family of simulations with these ratios but with different masses and momenta are compatible with changes in the total masses of the injected gas within factors of about ten of the highly uncertain mass and momentum estimates for $^{12}$CO in OH231 (e.g., A+01, B+02).

The initial conditions that we used for the simulations (Table 1) provide important glimpses into all but perhaps the first century of the nebula's evolution. The similar initial densities and kinematics of the pairs of lobes and sprays suggest that they were formed in a common event. This result is a natural outcome; it was never forced on the models (except as a first guess for the initial conditions at the start of this program). The shaping of the present structures of OH231 occurred early at times when the density of the flows was less than or comparable to that of the ambient gas. Once the ensemble of flows emerge into the outer, low-density zones of the ambient AGB winds they expand ballistically, as expected for "heavy" or "fast-propagating" flows (Soker 2002ApJ...568..726S).

While the construction of a successful hydro model for OH231 is of interest, the lasting value of this work is really the constraints that the model imposes on the history of the original mass ejection process. Before going further let us note that the ejection speeds are fixed by measurements, so flow momenta scale with flow masses. Then from Table 1, we find masses and momenta of the southern spine (bulb) derived from the best-fit model are 0.1 $M_\odot$ and $8 \times 10^{39}$ gm cm s$^{-1}$ (2.3 $M_\odot$ and $1.8 \times 10^{41}$ gm cm s$^{-1}$), respectively. The corresponding values for the northern lobe are 0.1 $M_\odot$ and $4 \times 10^{39}$ gm cm s$^{-1}$ (0.5 $M_\odot$ and $0.2 \times 10^{40}$ gm cm s$^{-1}$), respectively. Note that the bulb (that is not directly detectable in $^{12}$CO) dominates these estimates. The total ejected mass and momentum are thus 3 $M_\odot$ and $2.1 \times 10^{41}$ gm cm s$^{-1}$, respectively. (In effect, they are anchored to the directly measured value of mass of the slow flow (i.e., the $^{12}$CO(1-0) flux of the narrow line), about which there is considerable observational uncertainty (section 2).) These values are exceptionally large among all other pPNe.

This paper is the first of several like it in which hydrodynamic models similar to the one developed here in order to fit detailed observations of other symmetric PNe. Our primary scientific goal in this series is to characterize the aspects of the histories that pPNe may share in common. To that end a detailed model for one pPNe (and a fairly unique one at that) is simply a first step. Hence our hydrodynamic studies of pPNe will continue on a case-by-case basis into the future.

*Acknowledgements:* B.B. wishes to recognize fifty years of boundless mentoring and loving friendship of his thesis adviser, Prof. Yervant Terzian.

We are grateful to J. Alcolea for useful early discussions. Nearly all of the hydrodynamic models in this paper were run in computers maintained by the Center for Integrated Research Computing of the University of Rochester. (A few of the very early models used the Extreme Science and Engineering Discovery Environment (XSEDE), which is supported by National Science Foundation grant number ACI-1053575.)

Many of the images used in this project were made possible by HST GO grant 11580. Support for GO11580 was provided by NASA through a grant from the Space Telescope Science Institute, which is operated by the Association of Universities for Research in Astronomy, Incorporated, under NASA contract NAS5-26555. Additional data that were used in this paper were obtained from the Multimission Archive (MAST) at the Space Telescope Science Institute (STScI). STScI is operated by the Association of Universities for Research in Astronomy, Inc., under NASA contract NAS5-26555. Support for MAST for non-HST data is provided by the NASA Office of Space Science via grant NAG5-7584 and by other grants and contracts.

*Facility:* HST (WFC3)

References

Alcolea, J., Bujarrabal, V., Sánchez Contreras, C., Neri, R., & Zweigle, J. 2001A&A...373..932A (A+01)
Balick, B. & Frank, A. 2002ARA&A..40..439B
Bujarrabal, V., Alcolea, J., Sánchez Contreras, C., & Sahai, R. 2002A&A...389..271B (B+02)
Bujarrabal, V., Castro-Carrizo, A., Alcolea, J., & Sánchez Contreras, C. 2001A&A...377..868B (B+01)
Chen, Z., Nordhaus, J., Frank, A., Blackman, E.G., & Balick, B. 2016MNRAS.460.4182C
Cunningham, A., Frank, A., & Hartmann, L. 2005ApJ...631.1010C
Cunningham, A.J.; Frank, A., Varnière, P., Mitran, S., & Jones, T.W. 2009ApJS..182..519C
De Marco, O. 2009PASP..121..316D
Forde, K.P. & Gledhill, T.M. 2012MNRAS.421L..49F
Huarte-Espinosa, M., Frank, A., Blackman, E.G., Ciardi, A., Hartigan, P., Lebedev, S.V., & Chittenden, J.P. 2012ApJ...757...66H
Iaconi, R., Reichardt, T., Staff, J., De Marco, O., Passy, J-C., Price, D., Wurster, J., &; Herwig, F. 2017MNRAS.464.4028I
Kastner, J.H., Weintraub, D.A., Zuckerman, B., Becklin, E.E., McLean, I., & Gatley, I., 1992ApJ...398..552K
Lee, C-F. & Sahai, R. 2003ApJ...586..319L (LS03)
Meakin, C.A., Bieging, J.H., Latter, W.B., et al. 2003ApJ...585..482M
Nordhaus, J., Blackman, E.G., & Frank, A. 2007MNRAS.376..599N
Riera, A., Raga, A.C., & Alcolea, J. 2005RMxAA..41..147R
Sánchez Contreras, C., Bujarrabal, V., & Alcolea, J.1997A&A...327..689S
Sánchez Contreras, C., Bujarrabal, V., Miranda, L.F., Fernández-Figueroa, M.J. 2000A&A...355.1103S (S+00A) optical
Sánchez Contreras, C., Gil de Paz, A., & Sahai, R. 2004ApJ...616..519S
Soker, N. 2002ApJ...568..726S
Soker, N. & Kashi, A. 2012ApJ...746..100S
Staff, J.E., De Marco, O., Madonald, D. et al., 2016MNRAS.458..832S